\documentclass[preprint,5p,twocolumn]{elsarticle}

\usepackage{lineno,hyperref}

\usepackage{epstopdf}

\usepackage[]{siunitx}

\journal{Nuclear Instruments and Methods in Physics Research  A}

\usepackage{xspace}
\newcommand{\PANDA}{$\overline{\textrm{P}}\textrm{ANDA}$\xspace}

\begin{document}

\begin{frontmatter}

\title{Investigation on intense axial magnetic field shielding with a large melt cast processed Bi-2212 tube}

\author[1]{Alaa Dbeyssi \corref{mycorrespondingauthor}}
\ead{adbeyssi@uni-mainz.de}

\author[1,2]{Bertold Froelich \fnref{myfootnote1}}
\ead{bertoldfroehlich@outlook.de}
\fntext[myfootnote1]{Currently at: Bundesanstalt f\"ur Strassenwesen, Br\"uderstrasse 53, 51427 Bergisch Gladbach, Germany}

\author[1,2]{Maria Carmen Mora Espi}

\author[1,2,3]{Frank Maas}

\author[1]{Luigi Capozza}

\author[1]{Oliver Noll}

\author[1]{Yadi Wang \fnref{myfootnote2}}
\fntext[myfootnote2]{Currently at: North China Electric Power University, Beijing 102206, People's Republic of China}

\author[1]{Dexu Lin \fnref{myfootnote3}}
\fntext[myfootnote3]{Currently at:  Institute of Modern Physics, Chinese Academy of Sciences, Nanchang Rd. 509, Lanzhou 730000, China}

\cortext[mycorrespondingauthor]{Corresponding author}

\address[1]{Helmholtz Institute Mainz, Staudinger Weg 18, D-55099 Mainz, Germany}
\address[2]{Institute of Nuclear Physics, Johannes Gutenberg University, D-55099 Mainz, Germany}
\address[3]{PRISMA + Cluster of Excellence, Johannes Gutenberg University, D-55099 Mainz, Germany}


\begin{abstract}
The feasibility of shielding axial magnetic fields up to 1.4~T, using a Bi-2212 hollow cylinder, is investigated at a temperature of 4.2~K. The  residual magnetic flux density  along the axis of the tube    is measured at external fields of 1~T and 1.4~T.  The shielding factor, defined as the ratio between the applied and the residual magnetic flux densities at the center of the tube, is measured to be $32\times 10^4$ at 1~T and  $12\times 10^3$ at 1.4~T. The induced current density is evaluated from the measurements taking  the thickness of the tube into account. The stability of the measurements over time is also addressed. Numerical simulations for  the external and the residual magnetic flux densities are performed and compared to the experimental results.   The study shows a high shielding performance of the Bi-2212 superconductor tube at 4.2~K up to 1.4~T. 

\end{abstract}


\end{frontmatter}


\section{Introduction}
\label{intro}

The understanding  of the nucleon structure from the QCD theory is one of the central issues in hadron physics. The structure of the nucleon can be described by structure functions  that are experimentally accessed though the measurements of electromagnetic processes in scattering and annihilation experiments.  In the last decade, the experimental techniques have been substantially and continuously improved yielding new insights in nucleon structure. A recent breakthrough is due to the implementation of  polarization techniques and the measurements of  spin dependent amplitudes of  electromagnetic processes. Among these techniques,  polarization experiments that make use of  the target polarization  have provided important and unique information on the structure of hadrons. Ambitious experimental programs are foreseen at the accelerator facilities using a transversely polarized targets with the aim to  achieve a complete and precise description of the hadron structure.  In particular, a transversely polarized target at the \PANDA experiment \cite{Lutz:2009ff,PANDA:2021ozp} is foreseen to measure polarization observables in antiproton-proton annihilation processes  \cite{Frohlich:2017ald}. Changing the spin alignment of the target in spectrometers employing strong magnetic fields such as in \PANDA, is a challenging task.  In order to operate a transversely polarized target within the \PANDA spectrometer, the longitudinal 2~T magnetic field of its solenoid has to be sufficiently shielded. The residual field at an applied external field of 2~T should be as low as possible, to maintain a strong degree of transverse polarization.  In addition, a homogenous residual field in a volume that covers the whole region of the target is required. Other important characteristics, needed to avoid spoiling the detection and  particle identification efficiency of the spectrometer, are the low material budget  introduced by the shielding material and a compact shielding volume of the magnetic field.

Tubular high temperature superconductors are promising solutions to meet the requirements for polarized target experiments~\cite{Denis2007,Fagnard2009}.  A superconducting  tube can be viewed as a a concatenation  of  single rings.  A variation of  an external magnetic flux in an ideal superconducting ring leads to a near to steady induced current that creates an opposing magnetic field  as a  consequence of the Faraday law of induction. The set of the rings in a tube is therefore able to induce an opposing magnetic field that follows the  distribution of the external one,  and to provide a homogeneously distributed residual field  in  a region sufficiently large for the installation of a polarized target. A superconductor tube of finite thickness is  not exactly  an ideal conductor, and a penetration of the external magnetic flux  above a certain threshold value can be observed.  For a given attenuation level, the induced current density of the shielding material  and the thickness of the tube determine the maximum magnetic field that can be shielded. They describe the maximum shielding currents that can flow in the shielding tube.

The shielding performance of  the high temperature superconductors such as  YBCO, Bi-2212, Pb-doped Bi-2223 and MgB2, using tubular samples and different manufacturing techniques,  has been experimentally tested in different studies~\cite{MILLER1993180,PLECHACEK199695,Cavallin_2006,Fournier1994,osti_1419740,Fagnard_2010}.   In the present work, the particular choice of Bi-2212 as a high temperature superconducting material is motivated  by the measurements of F.~Fagnard {\it et al.} \cite{Fagnard_2010}. A large melt cast    Bi$_2$Sr$_2$CaCu$_2$O$_8$  (Bi-2212) hollow cylinder of 80 mm length, 8 mm inner radius and 5 mm wall thickness has been studied in axial magnetic fields at high temperatures above 10~K ~\cite{Fagnard_2010}.  At  10~K,  a record value of  1~T magnetic field is shielded with a shielding factor of $10^3$~\cite{Fagnard_2010}. This corresponds to an induced current density of 16  kA/cm$^2$.  The induced current density in a high-temperature superconductor  decreases with rising the temperature.  A better performance  below 10~K is therefore expected.  A Bi-2212  tube is a suitable shielding candidate for polarized target experiments, due to its high transition temperature (92~K),  and low density  (6.3 g/cm$^3$) and low  average-$Z$ which minimize the energy loss of the reaction products and less affect their momentum measurement.

In this work, we study the feasibility of shielding intense axial magnetic fields using a high temperature  superconducting Bi-2212 tube at a temperature of 4.2 K. A sample of a  melt cast  Bi$_2$Sr$_2$CaCu$_2$O$_8$  hollow cylinder  from  Nexans\footnote{Nexans SuperConductors GmbH, D-50351 H\"urth, Germany}  with geometrical characteristics in line with the setups of modern high energy physics experiments is used. The dimensions of the tube are listed in Tab.~\ref{tab:geotube}. The current density that can be carried by a superconductor material and by consequence its shielding performance depends strongly on the fabrication process. Details on the melt cast process and the centrifugal technique used in the manufacturing of the tube used in this study are discussed in Ref.~\cite{Bock1995}.   

 The shielding measurements are carried out using a dedicated apparatus consisting of a cryostat filled with liquid helium, two superconducting magnets, and a Hall probe. The setup of the experiment and the  results of the measurements are presented in section~\ref{sec:experiment}. Numerical simulations are developed and compared to the measurements.  They are described in section~\ref{sec:numerical}. These investigations help in the development of shielding prototypes that fit the specific requirements of the future polarized target experiments.

\begin{table}[h]
	\begin{tabular}{c c c}
	\hline
        Length & $L_{st}$ & 150 mm \\
        		Outer radius & $r_{st}$ & 25 mm \\
		 Wall thickness&$d$& 3.5 mm\\
		\hline
	\end{tabular} 
	\caption{The geometrical characteristics of the Bi-2212 shielding tube. }
	\label{tab:geotube}
\end{table}


\section{Experiment}
\label{sec:experiment}
In this section, the experimental measurements for testing the shielding efficiency of the Bi-2212 tube are described.  A longitudinal magnetic field parallel to the wall of the shielding tube is applied, and measurements of the residual field along the axis of the tube are performed at a temperature of 4.2~K. The maximum magnetic field generated at the center is  1.4~T. The longitudinal component of the residual field is measured with a Hall probe placed on the axis of the tube with a manually driven moving system. In addition,  stability measurements over several days are taking place at constant external fields. The measurements are carried out in a liquid helium environment. In the first part of the experiment, the external magnetic field is measured  without the installation of the shielding tube.
\subsection{Setup}
\label{experimentalsetup}
\begin{figure*}[!htbp]
\centering
\includegraphics[scale=0.3]{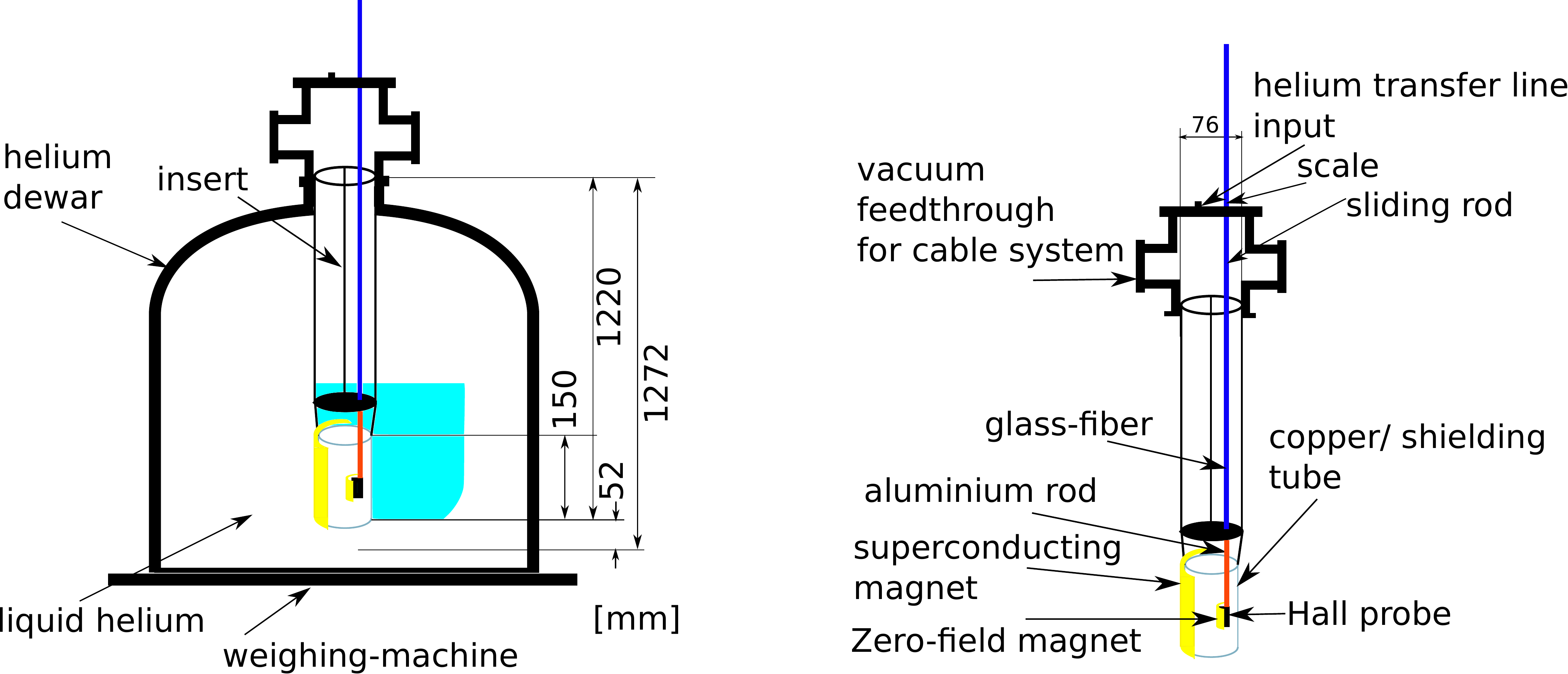}
	\caption{Illustration of the experimental setup: (left) dewar filled with liquid helium to keep the external magnet and the shielding tube at a temperature of \mbox{4.2 K}, and  (right) the insert with the shielding/copper tube and the external magnet. The Hall probe and the Zero-field magnet are mounted on the lower end of a sliding rod (glass fiber rod in the upper section of the insert is connected to an aluminum rod in the lower section).  The rod is equipped with a scale  to determine the position of the Hall probe.  The top of the dewar  is equipped with bores that are used to lead the cable system out from inside. The helium transfer line, used to fill the dewar, can be also attached to the system.}
	\label{dewar}
\end{figure*}

\begin{figure}[h]
\centering
\includegraphics[scale=0.05]{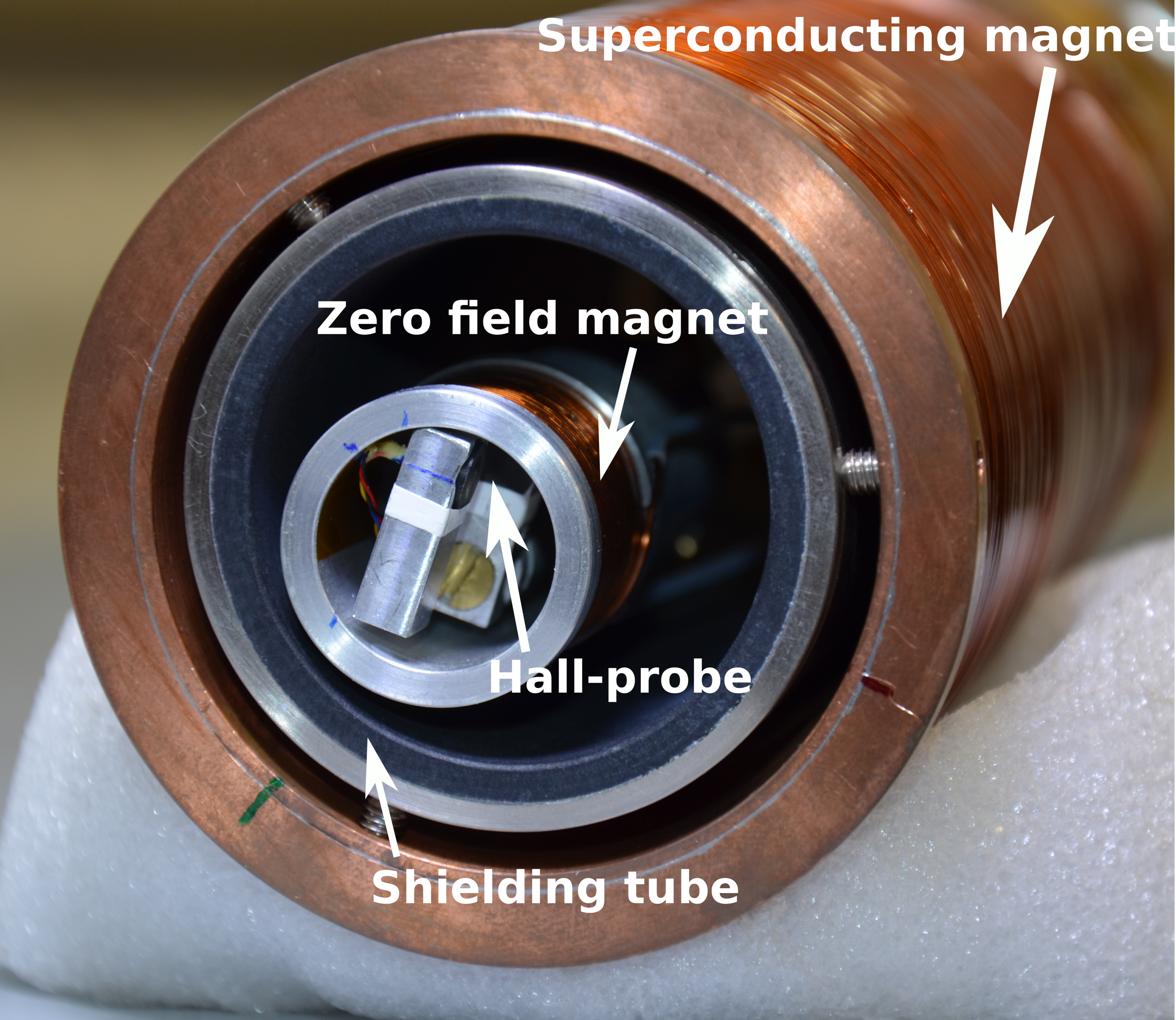}
	\caption{The external superconducting magnet with the shielding tube.  The Hall probe is placed on a stick that is moved to its position on the axis of the Bi-2212 tube. In the first part of the experiment, a copper tube  is replacing the shielding tube.}
	\label{tubesystem2}
\end{figure}

The setup of the experiment is illustrated in Fig.~\ref{dewar}. A tube system (Fig.~\ref{tubesystem2}) consisting of an external magnet, the shielding tube, a Hall probe and an internal magnet (called hereafter a Zero-field magnet) is placed in a dewar filled with  liquid helium.  An insert is used to hold the tube system and to guide the leads of the dewar. The level of the helium is controlled during the experiment thought the measurements  of the dewar weight.  A pressure sensor is mounted on the top of the inner part of the dewar. All components used in this experiment are made of non-ferromagnetic materials.
\subsubsection{External magnet}
The external magnet was designed and constructed to provide intense magnetic fields of at least 2~T. The size parameters of the external magnet are listed in Tab.~\ref{tab:geomag}. The wires are made of multi-filamentary Niobium-Titanium (NbTi) with a transition temperature around 9~K. The magnetic field at the center of the external magnet is measured to be   $833.6\pm1.6$~G   (1~G=$10^{-4}$~T)  at 1~A. The magnet quenches at $24150\pm90$~G. A bipolar high current supply {\it Kepco BOP 1000W} is used to operate the external magnet.
\begin{table}[h]
\begin{tabular}{c c c}
\hline
Length & $L_{em}$ &  138 mm \\
Radius & $r_{em}$ & 31.5 mm \\
Number of windings per layer  & $N_{em}$ & 460 \\
Number of layers  & $n_{0}$ & 22 \\
Distance between windings & $d_w$ & 0.3 mm\\
\hline
\end{tabular}
\caption{The geometrical characteristics of the external magnet ($em$).}
\label{tab:geomag}
\end{table}
\subsubsection{Bi-2212 shielding/Copper tube}
The shielding tests are performed with the  Bi-2212 tube  described in Tab~\ref{tab:geotube}. It is held by a CuNiMn (LV7) tube with an inner radius of 25~mm and a wall thickness of 2~mm. The tube system is connected to the insert by a copper thread soldered to the tube. In the first part of the experiment, the  shielding tube is replaced by a copper tube  to measure the external magnetic field. 
\subsubsection{Hall probe}
The longitudinal component of the magnetic fields are measured using a Hall probe from Lake Shore (HGCA 3020).  The Hall probe is designed to operate in the temperature range  from 1.5~K  to 375~K.   Corrections to the linear response of the Hall probe, calibrated by Lake Shore, are applied. The accuracy of the Hall probe operated at a nominal Hall probe control current of  100~mA is better than 0.1$\%$  up to 2~T.   The devices Digistant~64256~T  from Burster Pr{\"{a}}zisionsmesstechnik  and Prema~5017 from PREMA Semiconductor  are used for the control current and the readout of the Hall probe voltage,  respectively. The Hall probe holder is fixed on a sliding rod that can be moved along the axis of the shielding tube. The position of the Hall probe is measured using a scale fixed on the top of the dewar.
\subsubsection{Zero-field magnet}
The Zero-field magnet is a normal conducting coil on an aluminum holder placed directly on the top of the Hall probe to ensure its functionality  during the shielding measurements. The geometrical parameters of the Zero-field magnet are given in Tab.~\ref{tab:param_zfm}. The current supply of the Zero-field magnet is provided by  the device Instek~PSP~603 from  GW INSTEK.
 \begin{table}
 	\centering
 	\begin{tabular}{c c}
 		\hline  
		  Length & 20 mm \\
		 Inner diameter& 22 mm \\
		 Number of winding per layer&  27\\
		 Number of layers&  2\\
		 Wire diameter&  0.75 mm\\ 
 	        \hline 
 	\end{tabular} 
 	\caption{The geometrical characteristics of the Zero-field magnet.}
 	\label{tab:param_zfm}
 \end{table}
The maximum field that can be generated  by the Zero-field magnet  is $\sim ~210 $~G.    At 1~A, the magnetic field at the center is 21.9~G. The magnetic field strength outside the Zero-field magnet  is very small and its induced current  in the shielding tube can be neglected.
\subsubsection{Data acquisition system}
A data acquisition system has been developed to control the power supplies of the magnets and the Hall probe. It  also used to read out the Hall probe and the pressure sensor voltages.  The readout and the control devices are connected to a computer running the EPICS software via a RS-232 interface. The time interval between the read-out cycles is 2 seconds. The data are synchronized with a time delay of less then one second, and they are written in the same output stream. 
\subsection{Description of the measurements}
\label{exp_proc}
 Using the setup described above, the following  measurements are carried out : 
\begin{itemize}
	\item[-] measurements of the the external magnetic flux density $B_{ext}$ without the Bi-2212 shielding tube: the response of the generated magnetic field to the applied current $I_{ext}$ is measured at the center of the external magnet.  In addition, $B_{ext}$ is measured along the axis of the external magnet  at fixed $I_{ext}$.
        \item[-]  measurements of the residual magnetic flux density $B_{res}$ at the center of the Bi-2212 shielding tube by increasing and decreasing $B_{ext}$ in the intervals $[-1,~1]$~T and $[-1.4,~1.4]$~T.
\item[-]  measurements of $B_{res}$ along the axis of the tube at fixed $B_{ext}$ (1~T and 1.4~T).
	\item[-]  measurements of $B_{res}$  with the Zero-field magnet:  in the presence of the shielding tube, a very low signal is expected to be detected by the Hall probe. The known magnetic field generated by the Zero-field magnet placed between the Hall probe and the shielding tube is used to test the response of the Hall probe. The current of the Zero-field magnet is increased gradually up to $~1$~A.  $B_{res}$ is determined from the linear fit  to the measured magnetic flux density as a function of the increased current. In the following, this value is called "inc." value. Another value,  called "dec.", is obtained in the same way by decreasing the current  of the Zero-field magnet. Before the start of the "inc." and "dec." measurements, the Zero-field magnet is turned off and two  additional data points, called "0 inc." and "0 dec." are collected, respectively.	
	\item[-]  stability measurements  over days  of $B_{res}$ at the center of the Bi-2212 tube,  at  fixed  values of  $B_{ext}$  (1~T and 1.4~T).
\end{itemize}

In each measurement series of $B_{res}$,  an offset defined as the measured  $B_{res}$  without applying any magnetic field, is determined experimentally and subtracted from the data.  
\subsection{Analysis of the measurements}
\label{resexp}
\subsubsection{Estimation of the uncertainty on the measurement of the residual magnetic field}
The uncertainty in the measurement of  $B_{res}$  is mainly given by the uncertainty on the readout of the Hall probe voltage. The fluctuations of the Hall probe voltage are studied by recording the output of the {\it{Prema 5017}} voltmeter for about 12 hours. The output is divided into six data sets corresponding each to two hours of data taking.  Each data set is fitted with a Gaussian function where its standard deviation $\sigma$ is determined. The largest value  $\sigma=0.14$ $\mu$V is taken as the statistical error of the measurements. At the helium temperature, this value corresponds to a statistical error of 0.17~G  on the measurement of $B_{res}$. In addition, the drift given by the manufacturer for a 24 h measurement is less than 0.6~$\mu$V (0.73~G).  The two uncertainties are summed in quadrature and a conservative estimate for the uncertainty on one single measurement of $B_{res}$,  0.75~G, is  considered in this analysis.
\subsubsection{Measurement of the external magnetic flux density}
\label{extcal}
\begin{figure}[h]
\centering
        \includegraphics[scale=0.6]{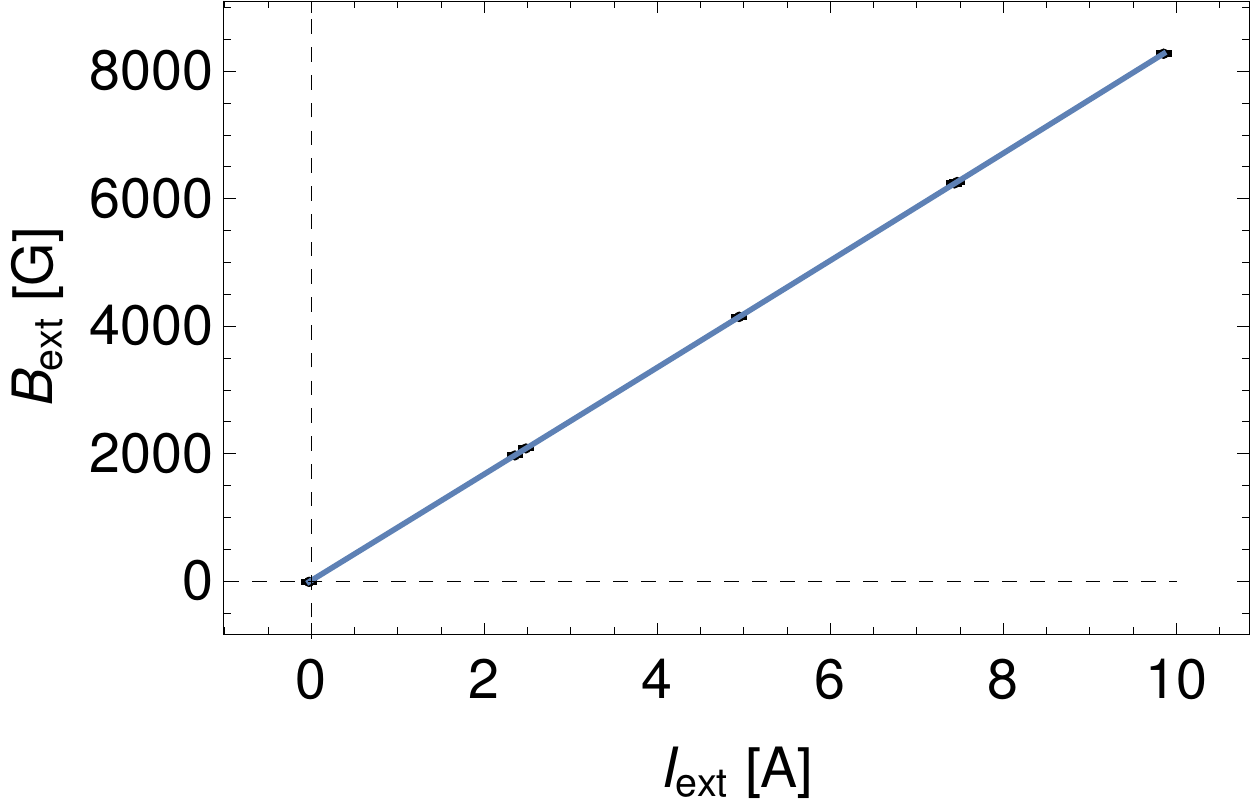}
       \caption{The measured values of  $B_{ext}$ as a function of  $I_{ext}$     (1~G=$10^{-4}$~T). The blue solid curve shows the fit to the data based on Eq.~\ref{fitpa}. The $\chi^2$/ndf of the fit is $0.2$.}
        \label{extfitplots1}
\end{figure}

\begin{figure}[h]
	\centering
\includegraphics[scale=0.6]{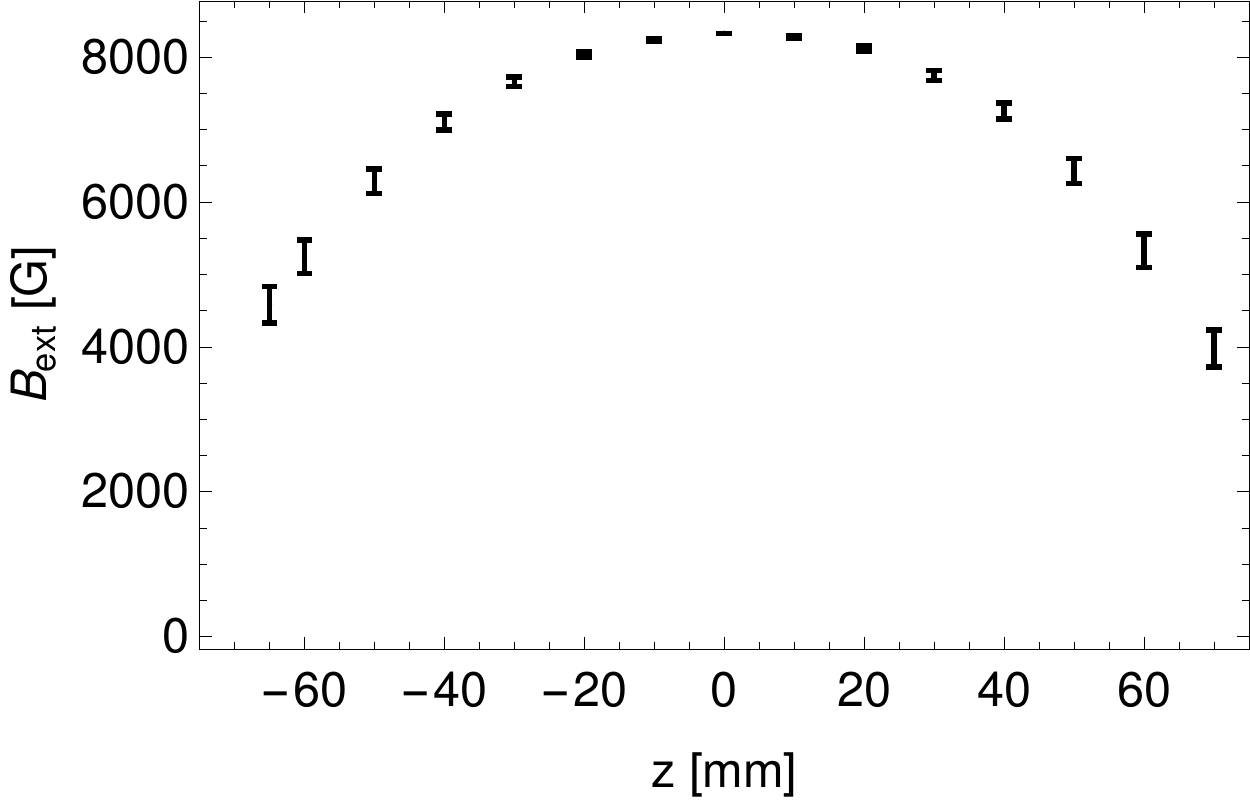} 
	\caption{The measured values   of $B_{ext}$  (1~G=$10^{-4}$~T) along the axis of the external magnet at $I_{ext}=(9.88\pm0.01)$~A.}
	\label{fig:ext-fieldmap}
\end{figure}

  A stable power supply current for the external magnet is obtained  by operating the {\it Kepco BOP 1000W} in a voltage mode. The  $B_{ext}$ is measured as the mean value of the data points collected with a stable current for at least 20~seconds. The same procedure is applied for the measurement of  $B_{res}$.  
The measured values of  $B_{ext}$ as a function of   $I_{ext}$ are shown in Fig.~\ref{extfitplots1}.  The data are well described by a second order polynomial function
 \begin{eqnarray}
 B_{ext} &=& a I_{ext}+bI_{ext}^2+c, \nonumber \\
 a &=&   833.03 \pm 1.17~\mbox{[G/A]}, \nonumber \\
 b &=&   0.58 \pm 0.14~\mbox{[G/A$^2$]}, \nonumber \\
 c &=&   9.42 \pm 1.65~\mbox{[G]}
\label{fitpa}
\end{eqnarray}
 where the parameters $a,~b$ and $c$ are determined from the fit to the data shown in Fig.~\ref{extfitplots1}. The uncertainty of the data is mainly determined  by the error in the setting of  $I_{ext}$ which is propagated to the uncertainty on $B_{ext}$.
 The measurements of  $B_{ext}$  along the axis of the  magnet are shown in Fig~\ref{fig:ext-fieldmap}. During these measurements, the maximum variation of  $I_{ext}$  is within $0.1\%$. The mean value is determined to be $I_{ext}=(9.88\pm0.01)$~A, and the  measured value of  $B_{ext}$ at the center of the magnet is \mbox{$(8280 \pm 10)$~G}. 
\subsubsection{Measurements of the residual magnetic flux density at 1~T}
\label{1T}
\begin{figure*}[!htbp]
\centering
         \includegraphics[scale=0.9]{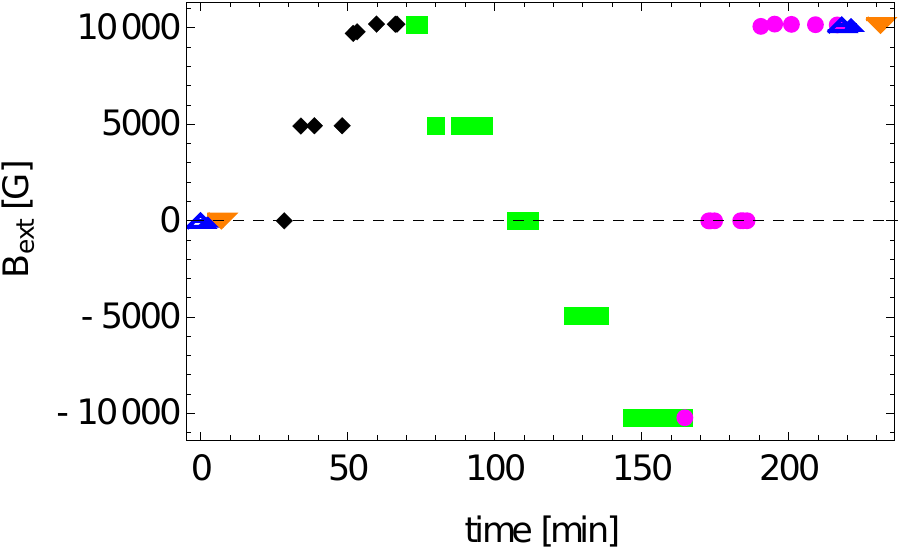} \hspace{0.1cm}
                     \includegraphics[scale=0.85]{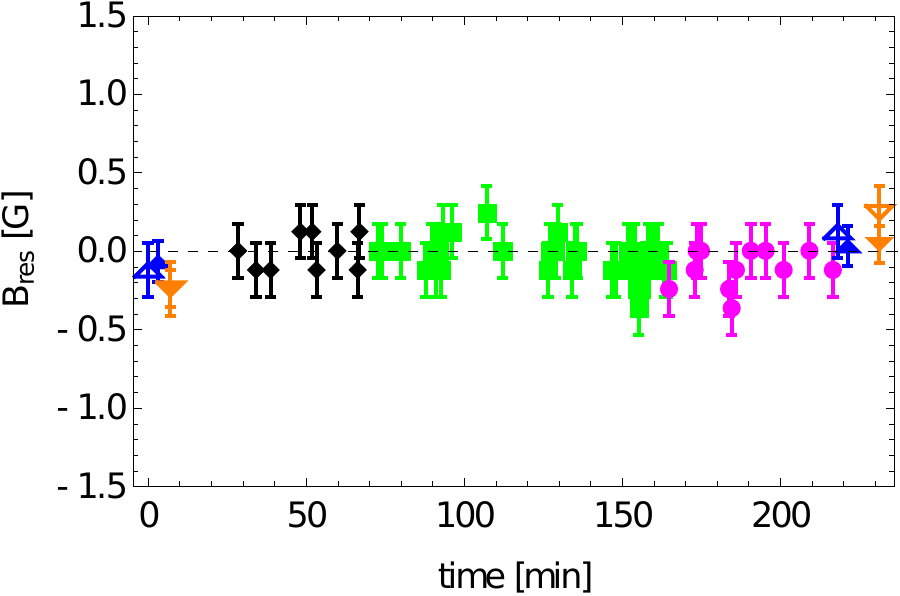} 
                     \caption{The values of   (left) $B_{ext}$ and (right)   $B_{res}$  (1~G=$10^{-4}$~T)  measured as a function of the time: $B_{ext}$ is 1) increased up to 1~T (black points); 2) decreased  down to -1~T (green squares); and  3)  increased again up to 1~T (magenta points).  The values of $B_{ext}$ are determined from Eq.~\ref{fitpa}.  The data obtained from the operation of the  Zero-field magnet are shown with blue filled triangles (inc. value), orange filled triangles (dec. value),   blue open triangles (0 inc. value),  and orange open triangles (0 dec. value).}
\label{Res1T}
\end{figure*}
\begin{figure}[h]
\centering
       \includegraphics[scale=0.6]{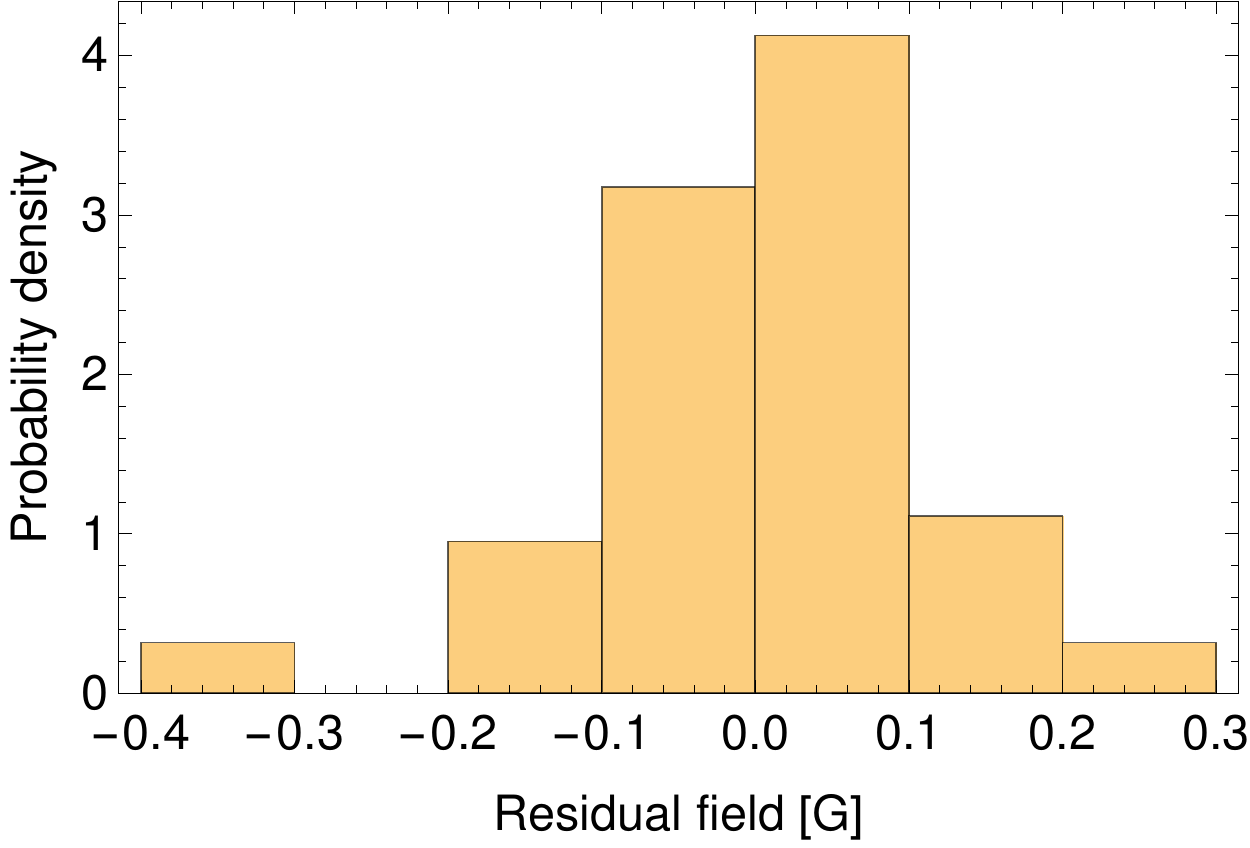} 
\caption{The values of $B_{res}$  (1~G=$10^{-4}$~T) after the offset subtraction ($\mu_{meas}$=-1~G), while increasing and decreasing the $B_{ext}$ to 1~T and -1~T, respectively.}
\label{fig:histo_bres}
\end{figure}
\begin{figure}[h]
\centering
       \includegraphics[scale=0.85]{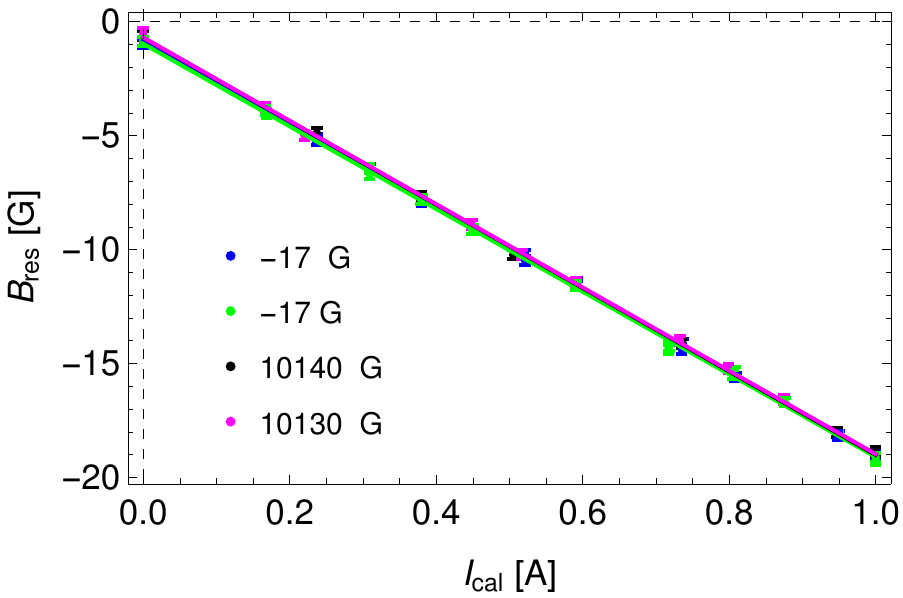}
       \caption{The measured values of  $B_{res}$   (1~G=$10^{-4}$~T)  as a function of $I_{cal}$ at  constant values of $B_{ext}$ (17~G and 10140~G): $I_{cal}$ is increased (blue and black points) and decreased  (green and magenta points).  The solid lines represent the linear fits to the data points.}
\label{figs:zfm_1T_1}
\end{figure}

The Bi-2212 is installed at the center of the external magnet and $B_{res}$ is measured  by the Hall probe while varying $B_{ext}$ between $\sim-1$~T and $\sim+1$~T.  The values of  $B_{res}$  measured at the center of the Bi-2212 shielding tube are shown in Fig.~\ref{Res1T}.  No increase of  $B_{res}$ is observed. The mean value of the  $B_{res}$ measurements,  $\mu_{meas}=-1$~G, is defined as the offset of these measurements. Figure~\ref{fig:histo_bres} shows the distribution of  $B_{res}$ after the offset subtraction. The statistical mean and fluctuation of this distribution are found to be  $\mu_{res}=B_{res}=0$~G and $\sigma_{res}=\Delta B_{res}=0.016$~G, respectively.
In the calculation of the $\mu_{meas}$ and $\sigma_{res}$ values, the data from the operation of the Zero-field magnet  are also included.  The linear fits to these data points, collected at $B_{ext}$=17~G  (stable measurements with zero external current are not feasible with the used power supply device) and 10140~G,  are shown in Fig.~\ref{figs:zfm_1T_1}.  The results of the fits on the determination of  $B_{res}$,  after the subtraction of the offset \mbox{-1 G}, are summarized in Tab.~\ref{table:zfm_1T_1}.  Its shown that no residual field is entering the shielding tube by increasing  $B_{ext}$  \mbox{up to 1 T}.

\begin{table}
\centering
\begin{tabular}{c c c c}
\hline
Mode & $B_{res}$   [G]& $B_{ext}$  [G]& $\chi^2\text{/ndf}$ \\ \hline 
  inc. & $0.2 \pm 0.13$ & $-17 \pm 2 $ &   0.1 \\
 dec. & $0.03 \pm 0.12$ & $-17 \pm 2 $ & 0.6 \\
 inc. & $0.3 \pm 0.13$ & $10140 \pm 14 $   & 0.5 \\
 dec. & $0.31 \pm 0.12$ &  $10130 \pm 14 $  & 0.5 \\
\hline
\end{tabular}
 \caption{The values of  $B_{res}$  (1~G=$10^{-4}$~T), after the the offset subtraction,  determined from the linear fits to the data collected with the Zero-field magnet (Fig.~\ref{figs:zfm_1T_1}). }
 \label{table:zfm_1T_1}
 \end{table}

The current density $J_{ind}$ of the Bi-2212 tube and its uncertainty are calculated  as
\begin{eqnarray}
J_{ind}&=&\frac{B_{ext}}{\mu_0 d},\nonumber \\
\sigma_{J}^2&=&\frac{\sigma_{ext}^2}{(\mu_0 d)^2}+\frac{B_{ext}^2}{(\mu_0 d^2)^2}\sigma_d^2
\end{eqnarray}
where $B_{ext}=(10140 \pm 14)$~G is the shielded external field, $d=3.5 \pm 0.2$~mm is the wall thickness of the tube, and $\mu_0=4 \pi \cdot 10^{-7}$ (Vs/Am).    The main source of  uncertainty in the determination of $J_{ind}= (23 \pm 2)\times 10^3$~(A/cm$^2$)  comes from the uncertainty on the measurement of the wall thickness $\sigma_d$ of the tube.  The shielding factor $SF$ is defined as
\begin{equation}
SF = \frac{B_{ext}}{B_{res}}.
\end{equation}
In the present case ($B_{res}=0$), the $SF$ is calculated using a MC simulation method.  A normal distribution for  $B_{ext}$  ($B_{res}$)  is  generated with a mean  and a  variance fixed respectively by the measured value  $B_{ext}=(10140 \pm 14)$~G  ($B_{res}= 0 \pm 0.016)$~G).  The two distributions for  $B_{ext}$ and  $B_{res}$  are used to calculate a probability density function for the estimation of the shielding factor.  At a confidence level of $95\%$, the shielding factor is determined to be $SF_{2\sigma}=3.2\cdot10^5$.

\begin{figure}[h]
  \centering
\includegraphics[scale=0.85]{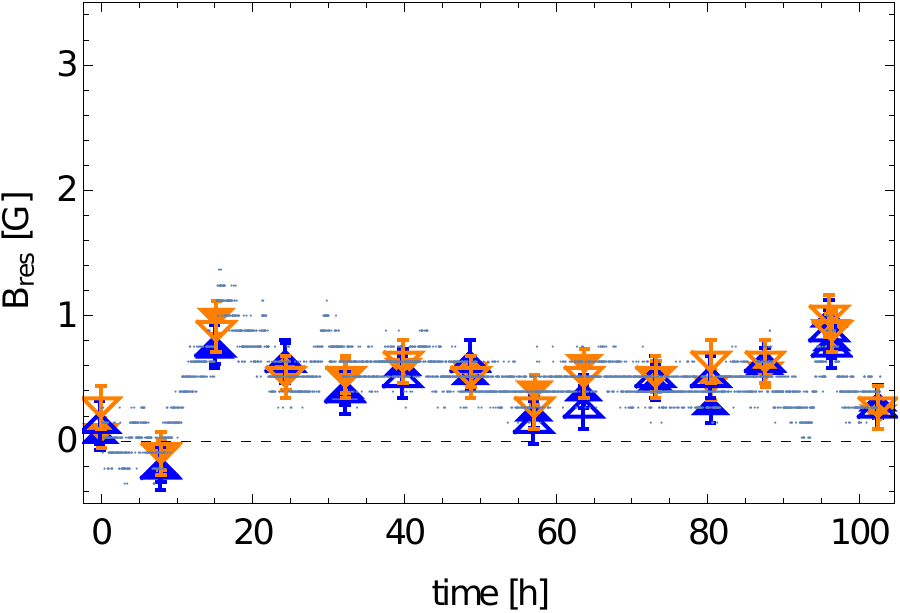}
\caption{Stability measurements of the residual field using the Bi-2212 shielding tube at  $B_{res}\sim1$~T. The total time of the measurements is four days. The data obtained from the operation of the  Zero-field magnet are shown with blue filled triangles (inc. value), orange filled triangles (dec. value),   blue open triangles (0 inc. value)  and orange open triangles (0 dec. value)}
\label{fig:stab1T}
\end{figure}

The stability of the residual field was measured at a constant external magnetic field of $~1$~T for about four days. The data are shown in Fig.~\ref{fig:stab1T}. An increase step of the residual field is observed after $\sim$ 9 hours  followed by a stable behavior. The mean values of the measured  $B_{res}$ and  $B_{ext}$  are summarized in Tab.~\ref{stab1T-offset}.

\begin{table}
\centering
\begin{tabular}{c c c}
	\hline
                                  &0 to 10 h&20 h to 100 h \\ \hline
	$B_{ext}$ [G] & 10081 $\pm$ 14 &  $(100 \pm 2)\times 10^3$   \\ 
	$B_{res}$ [G] & 0.000  $\pm$  0.008 & 0.499 $\pm$  0.003   \\ \hline	
\end{tabular}
\caption{The mean values of  $B_{res}$ (after the offset   ($-0.751 \pm 0.008$)~G subtraction)  and $B_{ext}$  (1~G=$10^{-4}$~T)   from the stability measurements (Fig.~\ref{fig:stab1T}). }
\label{stab1T-offset}
\end{table}

\begin{figure}[h]
	\centering
    \includegraphics[scale=0.95]{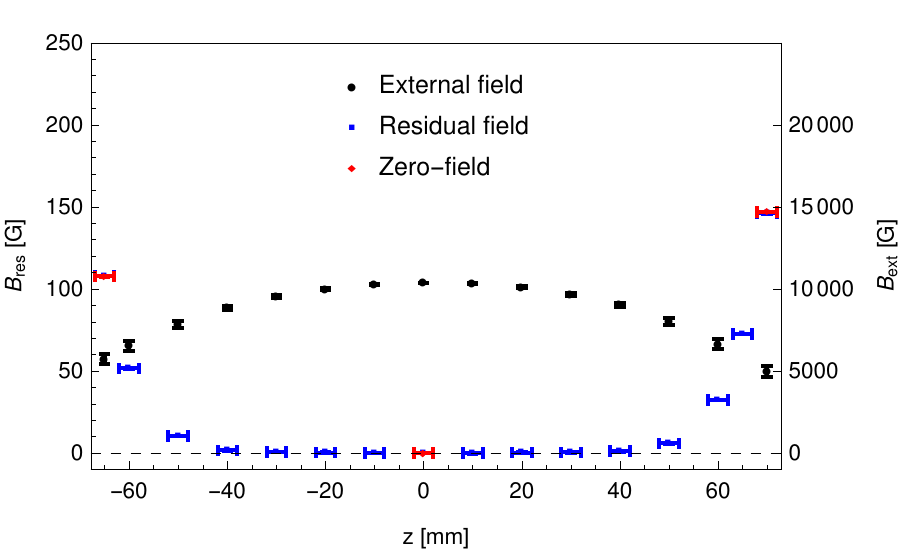} 
	\caption{The distribution of  $B_{res}$   (1~G=$10^{-4}$~T) measured along the axis of the Bi-2212 tube (blue squares) .   The black points show the distribution of  $B_{ext}$ ($\sim$1~T at the center). The red diamonds  are the values of  $B_{res}$  determined from the operation of the Zero-field magnet.}
	\label{fig:plotfmap}
\end{figure}

The residual field along the axis of the shielding tube is measured at an external field  at the center of the tube  equal to  \mbox{$(10330\pm 14)$ G}.  The results of the measurements are shown in Fig.~\ref{fig:plotfmap}. A shielded length, where $B_{res}$ is lower than \mbox{1 G},  of  \mbox{80 mm} is observed from a  total length of \mbox{150 mm}.
\subsubsection{Measurements of the residual field at 1.4~T}
\begin{figure}[h]
	\centering
     \includegraphics[scale=0.85]{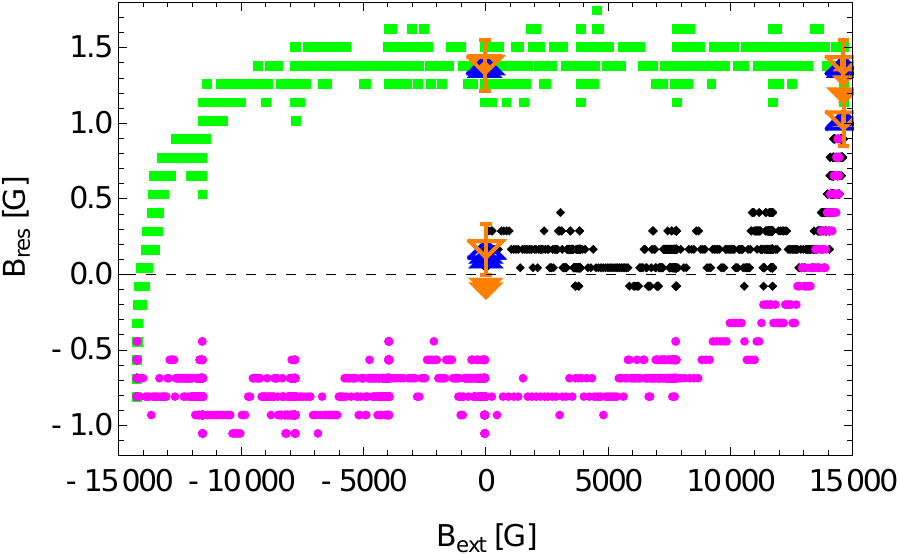} 
\caption{The measured values of   $B_{res}$  as a function of $B_{exp}$  (1~G=$10^{-4}$~T):  $B_{ext}$ is 1) increased up to 1.4~T (black points); 2) decreased  down to -1.4~T (green squares); 3) and increased again up to 1.4~T (magenta points).  The values of $B_{ext}$ are calculated using Eq.~\ref{fitpa}.   The data obtained from the operation of the  Zero-field magnet are shown with blue filled triangles (inc. value), orange filled triangles (dec. value),   blue open triangles (0 inc. value) and  orange open triangles (0 dec. value)}    
	\label{fig:plotbres}
\end{figure}

In this part of  the experiment,   $B_{ext}$ is varying between -1.4~T and 1.4~T and  $B_{res}$  is measured  at the center of the Bi-2212 tube, as shown in Fig.~\ref{fig:plotbres}. The data include also the measurements performed using the Zero-field magnet as described in the previous section. The results of these measurements are listed in Tab.~\ref{tab:ZFM_param_1T4}.  A slight increase of  the $B_{res}$ values, compared to the measurements performed at 1~T, is observed.  

The residual field is calculated as follows
\begin{equation}
B_{res}=B_{res}^{max}-B_{res}^{0},
\end{equation}
where $B_{res}^{max}$ and $B_{res}^{0}$ are the values of the residual field at \mbox{$B_{ext} = (14640 \pm 30)$ G} and $B_{ext}=(17 \pm 2)$~G, respectively. The residual fields $B_{res}^{max}$ and $B_{res}^{0}$ are determined from the average over the Zero-field magnet measurements. The results are given in Tab.~\ref{tab:ZFM_param_1T4}. 
\begin{table}
	\[\begin{array}{c c c c }
	\hline
	\text{time [min]}    & B_{res}  \text{[G]}     & B_{ext}   \text{[G]}  & \chi^2\text{/ndf} \\
	\hline
	0  ~\text{(inc)} & -0.22 \pm 0.13 & \text{17 } \pm \text{2 }  & 0.2 \\
	6  ~\text{(dec)} & -0.43 \pm 0.12 & \text{17 } \pm \text{2 }   & 0.9 \\
	10 ~ \text{(inc)} & -0.19 \pm 0.13 & \text{17 } \pm \text{2 }  & 0.2 \\
	14  ~\text{(dec)} & -0.4 \pm 0.12 & \text{17 } \pm \text{2 } 	& 1.1 \\
	67  ~\text{(inc)} & 0.72 \pm 0.13 & \text{14640 } \pm \text{30 } & 0.4 \\
	72  ~\text{(dec)} & 0.86 \pm 0.12 & \text{14640 } \pm \text{30 }  & 0.3 \\
	77  ~\text{(inc)} & 1.04 \pm 0.13 & \text{14630 } \pm \text{30 }  & 1.3 \\
	83  ~\text{(dec)} & 0.98 \pm 0.12 & \text{14630 } \pm \text{30 }  & 0.4 \\
        \hline
	\end{array}\]
	\caption{The values of  $B_{res}$ (1~G=$10^{-4}$~T), after the the offset ($-1.44\pm 0.04$)~G  subtraction,  determined from the linear fits to the data collected with the Zero-field magnet at   $B_{ext}$=17~G  and 14640~G.   The measurements show a penetration of the external field into the shielding tube  1.4~T.}
	\label{tab:ZFM_param_1T4}
\end{table}

 The shielding factor  and the  current density at   $1.4$~T are determined to be $SF = \frac{B_{ext}}{B_{res}}=122\times 10^2$ and $J_{ind}  = (33 \pm  3) \times 10^3$  A/cm$^2$ (Tab.~\ref{ext-res_1T4}), respectively.
 \begin{table}
 	\centering
 	\begin{tabular}{c c}
 		\hline 
                  $B_{ext}$ [G] & 14640 $\pm$  30  \\
 		$B_{res}$ [G] & 1.22 $\pm$ 0.06  \\
 		$SF$ &  $(12 \pm 1) \times 10^3$  \\ 
 	        $J_{ind}$ [A/cm$^2$] & $(33 \pm  3) \times 10^3$ \\ \hline
 	 	\end{tabular} 
 	\caption{The  values of   $B_{ext}$,  $B_{res}$ (1~G=$10^{-4}$~T),  $SF$  and $J_{ind}$  measured at the center of the Bi-2212 tube.  The offset (-1.44$\pm $0.04) is subtracted from the data.}
 	\label{ext-res_1T4}
 \end{table} 

\begin{figure}[h]
  \centering
  \includegraphics[scale=0.75]{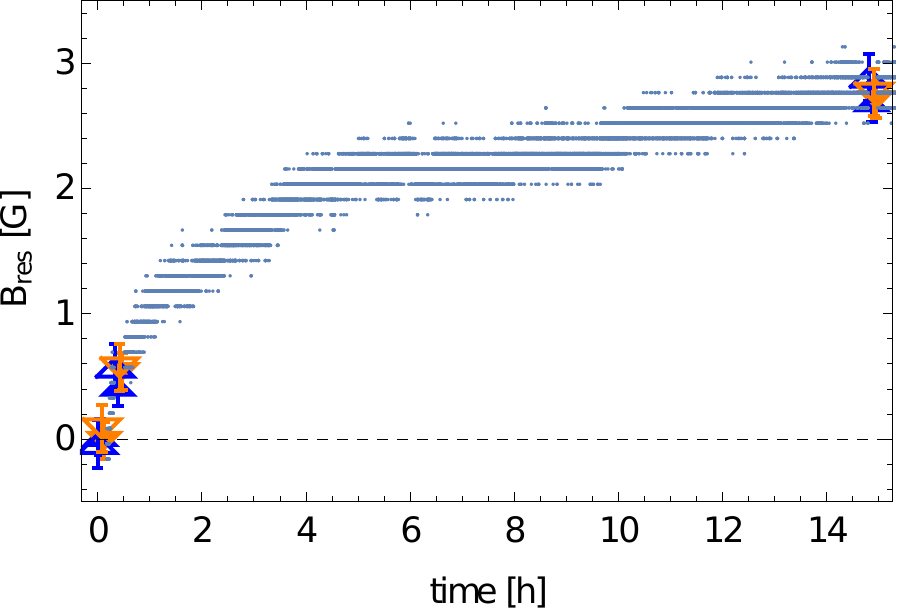} 
\caption{The measured values of  $B_{res}$ (1~G=$10^{-4}$~T) at the center of  the shielding tube at a constant    $B_{ext}$ of  $\sim$~1.4 T.  The  total time of  the measurements is $\sim $~14 hours.  Symbols are as in Fig.~\ref{fig:stab1T}. The offset \mbox{$(1.14 \pm 0.08)$ G} is subtracted from the data.}
\label{fig:stab1T4}
\end{figure}

\begin{figure}[h]
  \centering
  \includegraphics[scale=0.85]{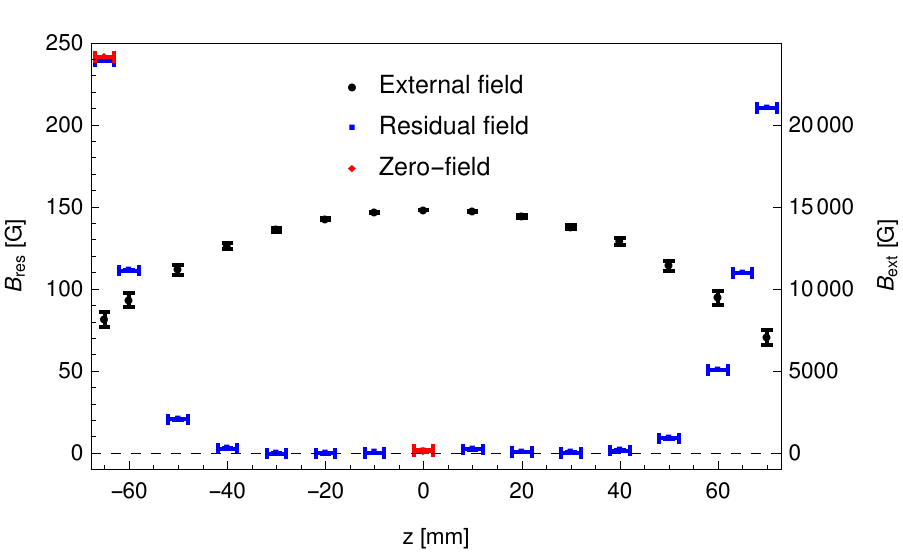} 
\caption{The distribution of  $B_{res}$ (1~G=$10^{-4}$~T)  measured along the axis of the Bi-2212 tube (blue squares) .   The black points show the distribution of  $B_{ext}$  ($\sim$1.4~T at the center). The red diamonds  are the values of  $B_{res}$  determined from the operation of the Zero-field magnet.}
\label{fig:fieldmap1T4}
\end{figure}

The stability of the residual field of the shielding tube at a constant external magnetic field of \mbox{$(14765 \pm 30)$ G } is measured for \mbox{14 hours} (Fig.~\ref{fig:stab1T4}). The  value  of $B_{res}$ is increased up to  $2.7  \pm  0.15$~G after 14~hours. 
The residual field along the axis of the tube is measured at constant  $B_{ext}=14793\pm 30$~G. The shielded length, with a residual field  less than \mbox{2 G},  is \mbox{80 mm} from a total  length of \mbox{150 mm}. The data are shown in Fig.~\ref{fig:fieldmap1T4}.
%
%
%
%
%
\section{Numerical simulations for shielding of a magnetic field  with a Bi-2212 tube}
\label{sec:numerical}

In addition to the experimental investigations on the shielding of  external magnetic fields with a  Bi-2212  tube, numerical simulations are performed and compared to the measurements. The simulations  will be used as a tool to  design a prototype of  geometrical characteristics that fits the requirements of a polarized target experiment with the  \PANDA spectrometer.

\begin{figure}[h]
\centering
\includegraphics[scale=0.46]{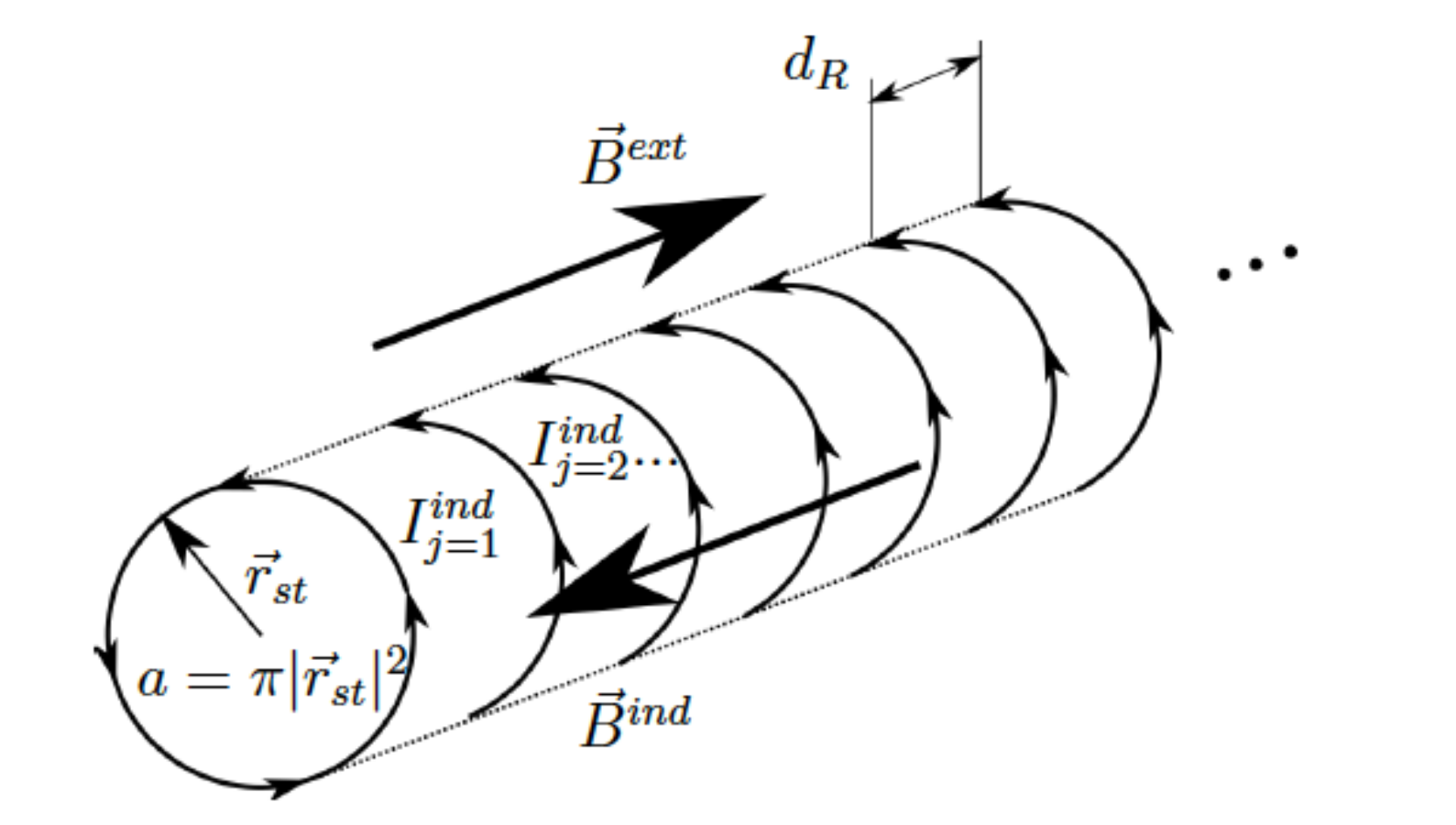}
    \caption{Schematic illustration of the discritization method applied on the shielding tube. The tube is divided in $N_{st}$ equidistant rings at distance $d_R$. Each of them is carrying a current $I^{ind}_j$, $j=1...N_{st}$ induced by the external field $\vec B^{ext}$ creating an induced magnetic field $\vec B^{ind}$.}
	\label{fig:current_loops}
\end{figure}

The induced current in a shielding tube which creates the shielding magnetic counter-field, depends on the position along the axis of the tube.  A simple Biot-Savart calculation with the assumption of a homogeneous current in the whole tube is not sufficient to calculate the residual magnetic field. Therefore, the induced current is calculated by using the exact forms of the Maxwell equations in integral form and the Biot-Savart law.  The method is based on the  discretization of the shielding tube  into rings \cite{FengG1985}  as shown in Fig.~\ref{fig:current_loops}. The tube is divided in $N_{st}=3750$ equidistant rings and the distance between the rings is $d_{R}=0.04$~mm.  The assumption of ideal circular conducting rings is used. Based on Faraday's law, the magnetic flux through the surface bounded by an ideal conductor is constant.  Assuming a zero residual magnetic flux ($F^{res}$) in the shielding tube at the beginning of the process, the external magnetic flux  ($F^{ext}$) is fully compensated by the induced flux ($F^{ind}$) inside the shielding tube ($F^{ind}=-F^{ext}$).  This  property  is used in the calculation of the induced current  and the induced magnetic flux density of the shielding tube.  $F^{ext}$ is calculated using the parameters of the external magnet. The simulations can be applied to any size of the shielding tube or the external magnet. In the following, the geometrical parameters for the tube and the magnet used in the experiment, are considered.

\subsection{Calculation of the external magnetic flux density}
\label{Biot-Savart}

\begin{figure}[h]
	\centering
\includegraphics[scale=0.16]{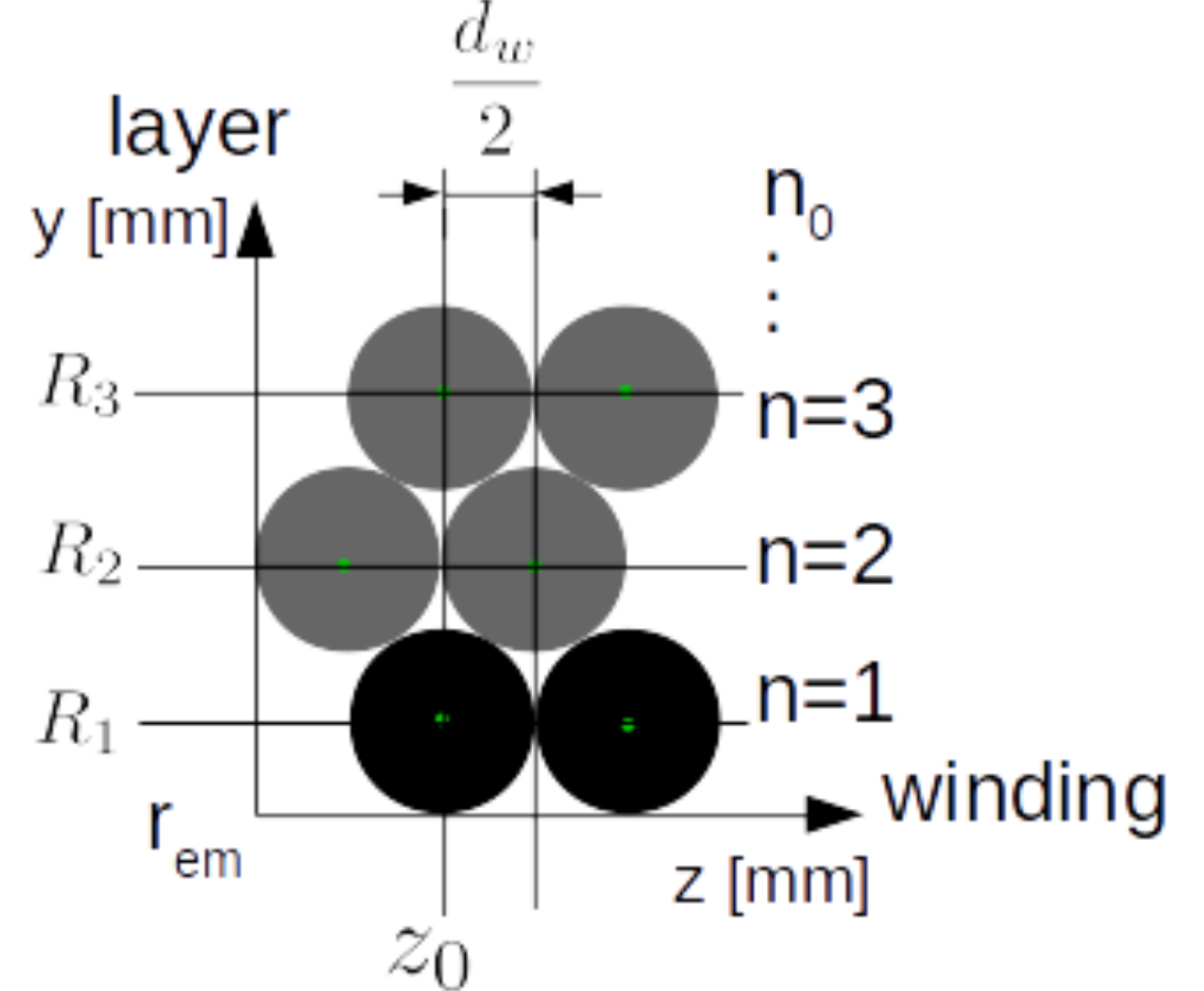}
	\caption{Example of the layout cross section of the external magnet with 2 windings and 3 layers.}
	\label{fig:layers}
\end{figure}

The external magnet used in the experiment has 460 windings and 22 layers ($n_0=22$).    An example for the layout cross section of 2 windings and 3 layers is shown in Fig.~\ref{fig:layers}. A winding of the external magnet has a simple geometry of a circular closed loop in approximation to a helix of a small height of one wire diameter compared to the total length of the external magnet. A reference system where the winding of the external magnet is located at the center of the the x-y plane, and creates a magnetic flux density in the x-z plane, is considered. The z component of the magnetic flux density at a position $P(r_x,~r_z)$ in the  x-z plane is  calculated via the Biot-Savart law \cite{Jackson} as follows:
\begin{eqnarray}
&&B^{w}_z(r_x,\Delta_z)=\frac{I^{ext}\mu_{0}}{4\pi} \\ \nonumber
&&\int_{0}^{2\pi}\frac{(R^{2}-Rr_x\cos\phi)}{(r_x^{2}-2r_xR\cos\phi+R^{2}+\Delta_z^{2})^{\frac{3}{2}}}d\phi,
\label{Biot-Savart_winding}
\end{eqnarray}
where $\Delta_z$ is the relative distance along the z-axis between the winding center and $P(r_x,r_z)$, $R$ is the radius of the winding,  and $\phi \in [0,2\pi ]$ describes the integration over the winding circle.  The $z$-component of the magnetic flux density of one winding is obtained by a summation over the  22 layers of the magnet:
\begin{eqnarray}
&&B^{wn}_z(r_x,\Delta_z) =  \frac{I^{ext}\mu_{0}}{4\pi} \sum^{n=n_0}_{n=1} \\ \nonumber
&&\int_{0}^{2\pi}\frac{({R_n}^{2}-R_n r_x\cos\phi)}{(r_x^{2}-2r_xR_n \cos\phi+{R_n}^{2}+(\Delta_z-z_n)^{2})^{\frac{3}{2}}}d\phi,
\end{eqnarray}
where the two variables $z_n$ and $R_n$ are functions of the layer number $n$  and describe the position of a single winding in the y-z plane:
\begin{eqnarray}
z_n&=&((n-1) \textrm{ mod } 2)\frac{d_w}{2}\\   \nonumber
R_n&=&r_{em}+\frac{d_w}{2}+(n-1)\sqrt{3}\frac{d_w}{2}
\end{eqnarray}
\subsection{Calculation of the external magnetic flux}
\begin{figure}[h]
\centering
\includegraphics[scale=0.15]{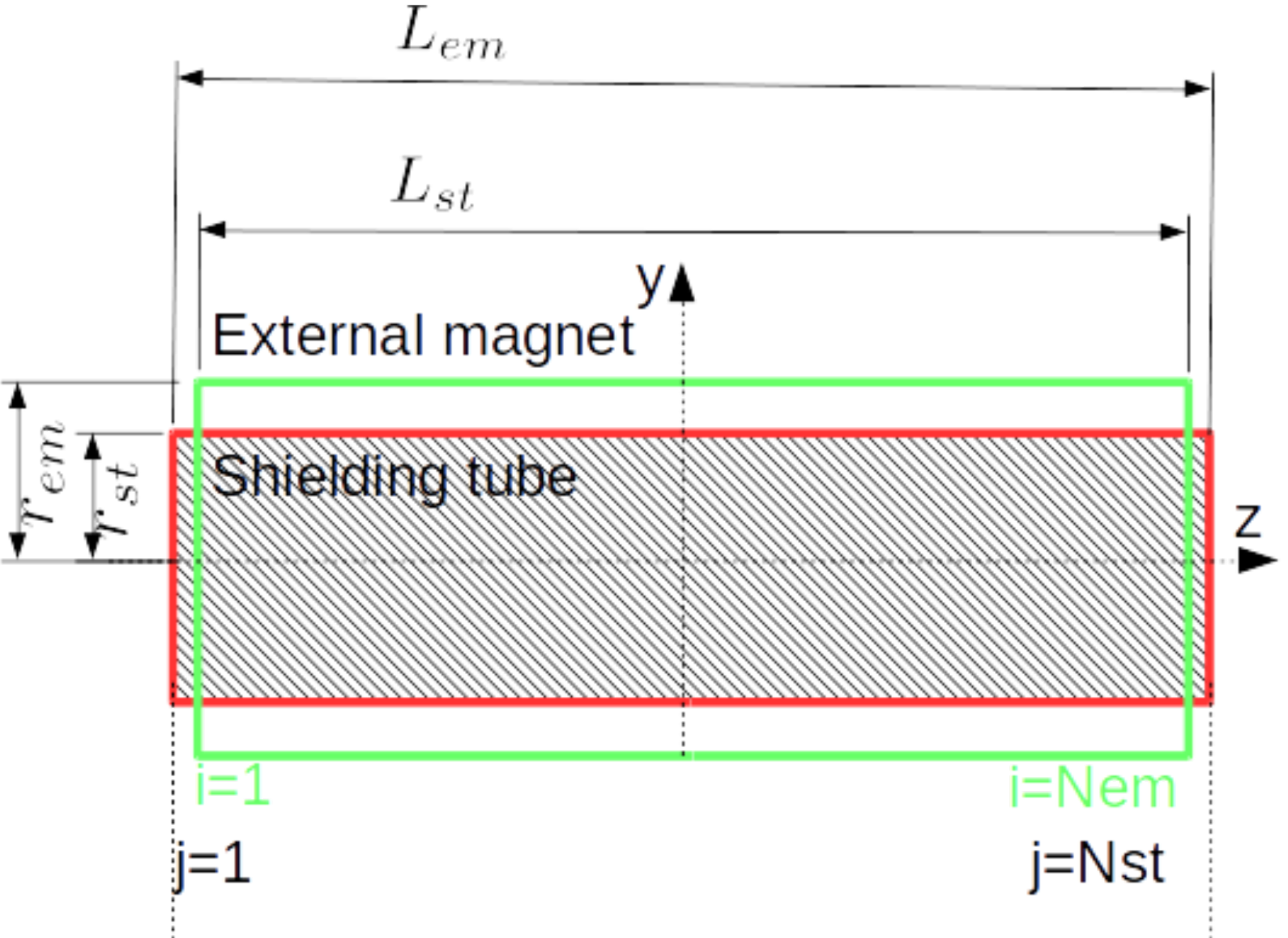}
\caption{Sketch of the geometry of the external magnet and shielding tube. The shaded area shows the region where the flux of the external magnet is calculated.}
\label{fig:fluxintube}
\end{figure}
\begin{figure}[h]
	\centering
\includegraphics[scale=0.6]{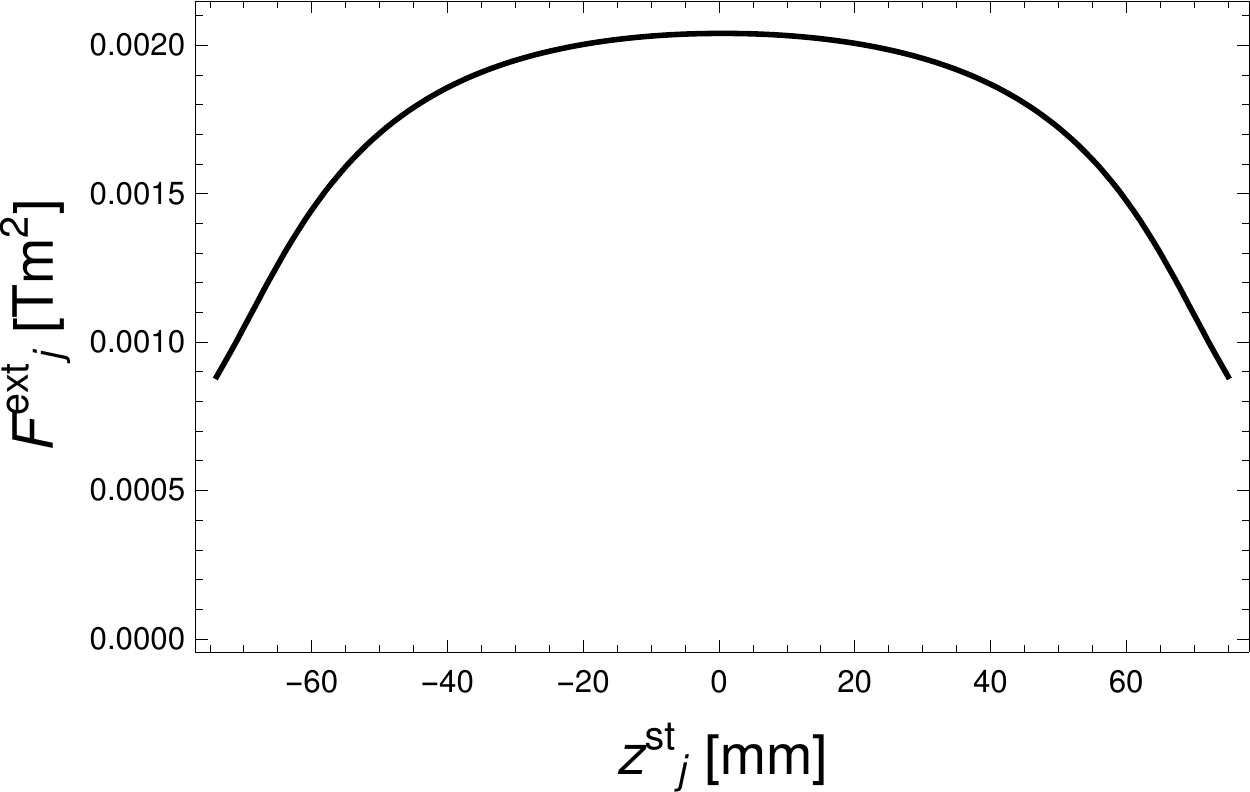}
	\caption{ The calculated magnetic flux of the external magnet in the shielding tube.}
	\label{fig:totflux}
\end{figure}

The external magnetic flux is calculated  based on the arrangement of the magnet and the shielding tube shown in Fig.~\ref{fig:fluxintube}. The external magnetic flux of one winding in a shielding tube ring is
\begin{equation}
F^{wn}(\Delta_z) = 2\pi\int_{0}^{r_{st}}B^{wn}_z(\rho,\Delta_z) \rho d\rho.
\end{equation}
The external flux in the shielding tube ring $j$ is calculated by the summation over all windings of the external magnet
\begin{equation}
F^{ext}_{j}=\sum_{i=1}^{N_{em}}F^{wn}(\Delta_{ij}),
\label{eq:flux}
\end{equation}
where $\Delta_{ij}=|z^{em}_i-z^{st}_j|$ is the distance  between the $z$ positions of the  the $i^{th}$ winding ($z^{em}_i,~i=1...N_{em}$) and the $j^{th}$ ring ($z^{st}_j,~j=1...N_{st}$).
Figure~\ref{fig:totflux} shows the external magnetic flux of the external magnet in the shielding tube as a function of $z^{st}_j$.  The value of $I^{ext}$ corresponds  to an external magnetic flux density at the center of the tube of  1~T.
\subsection{Calculation of the induced magnetic flux density}

\begin{figure}[h]
	\centering
\includegraphics[scale=0.4]{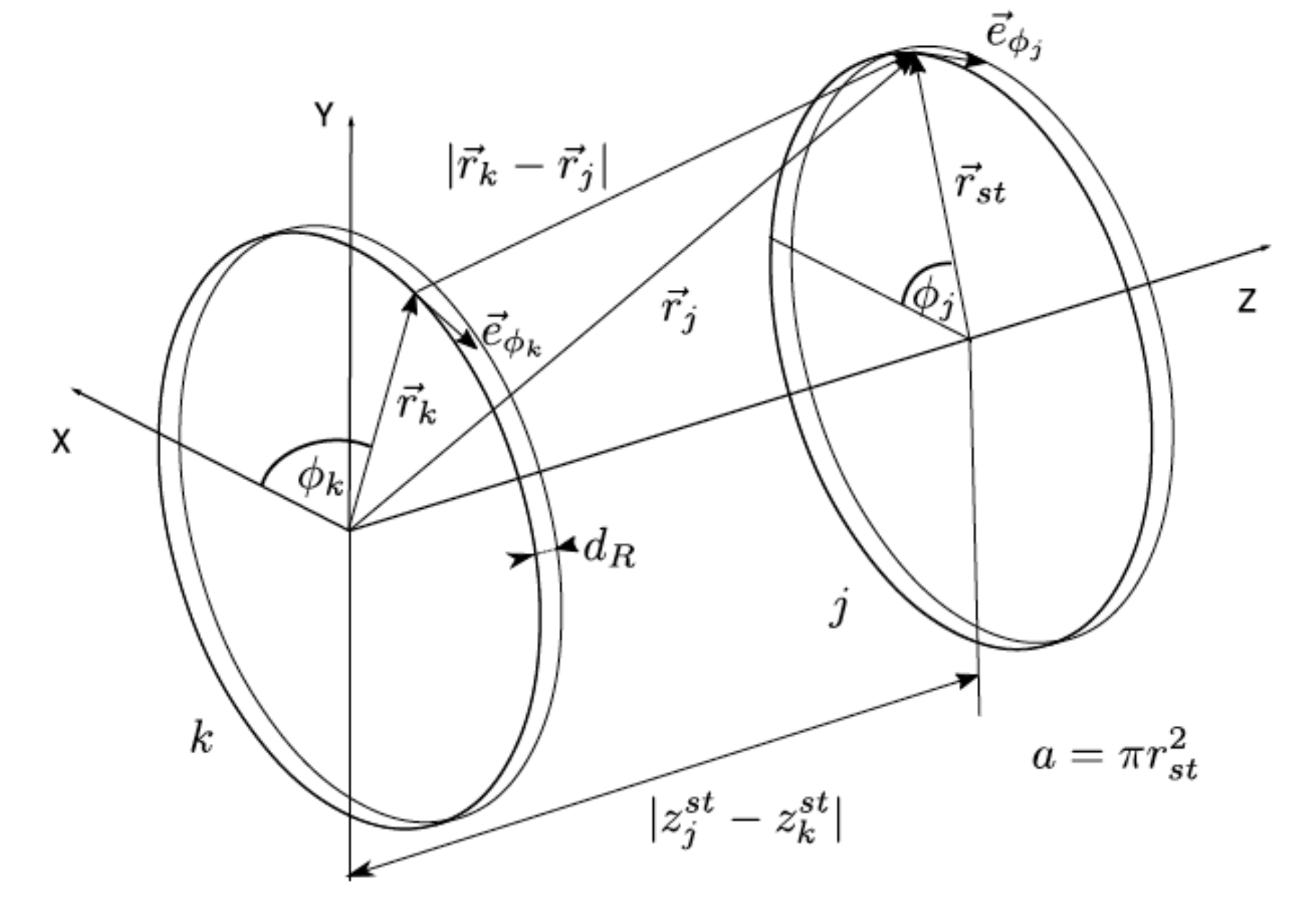}
	\caption{Sketch of two rings with symbols used to calculate the inductance matrix elements.}
	\label{fig:indelem}
\end{figure}

The derivation of the induced current for a system of thin wires can be found in Ref.~\cite{Jackson}. The calculations are extended here taking into account the width of the rings in the $z$-direction.  The flux in the ring $j$ of the shielding tube, $F^{ind}_j$, arising from a ring $k$, can be expressed  as a function the potential vector  $\vec{A_k}$ and the cylindrical coordinates shown in Fig.~\ref{fig:indelem}, as follows:
\begin{eqnarray}
\vec{A_k}(\phi_j,~z_j)&=&\frac{\mu_0 r_{st}}{4\pi}    \int_{0}^{2 \pi}   d\phi_k     \int_{z_k-\frac{d_R}{2}}^{z_k+\frac{d_R}{2}} d z'_k         \frac{{\vec j_k}(\phi_{k})}{|\vec r_k-\vec r_j |}, \nonumber  \\
{\vec j_k}(\phi_{k})&=&j_k \vec e_{\phi}(\phi_k), j_k=\frac{I^{ind}_k}{d_R},
\label{potvec}
\end{eqnarray}
where $\vec r_{k}$   and $\vec r_{j}$  are the vectors pointing to  the $k^{th}$ and $j^{th}$ rings, respectively.   The integration is performed  over the volume of the $k^{th}$ ring with radius $r_{st}$ and width $d_R$.
The induced magnetic flux $F'_j$ in the $j^{th}$ ring at position $z_j$ generated by the current in the $k^{th}$ ring is
\begin{equation}
F^{'}_j(z'_j)=\int \vec B_k\cdot da  = r_{st}\int_{0}^{2 \pi} \vec{A_k}\cdot \vec e_{\phi}(\phi_j){d\phi_j},
\label{flux}
\end{equation}
where $\vec B_k$ is the magnetic flux density created by the ring $k$. The flux in a ring $F^{ind}_j$ can be expressed as an average over the flux at a given $z$, $F^{'}_j(z_j)$, in the interval of the width $d_R$ of the $j^{th}$ ring. 
Using Eqs.~\ref{potvec} and ~\ref{flux},   an inductance matrix element $M_{kj}$ can be defined as follows: 
\begin{eqnarray}
F_j^{ind}&=&M_{kj}I^{ind}_k=-F_j^{ext},  \\ \nonumber
M_{kj}&=&  \frac{\mu_0   r_{st}^2}{4\pi d_R^2}
 \int_{z_j-\frac{d_R}{2}}^{z_j+\frac{d_R}{2}} dz'_j
\int_{0}^{2 \pi}  d\phi_j  \\ \nonumber
&&\int_{0}^{2 \pi}   d\phi_k
\int_{z_k-\frac{d_R}{2}}^{z_k+\frac{d_R}{2}}  d z'_k  
\frac{\cos(\phi_k-\phi_j)}{|\vec r_k-\vec r_j|}
\label{eq:indmateq}
\end{eqnarray}
where a summation over $k$ ($k$=1,..., $N_{st}$) expresses the contributions of all the rings  to the flux in the $j^{th}$ ring. The inductance matrix element depends only on the geometry of the rings, in the present case,  the radius $r_{st}$ and the  ring width $d_R$.  For the diagonal elements of the inductance matrix  ($k=j$),  a homogenous distribution of the current  over the thickness of the  ring is assumed \cite{BoS}:
\begin{equation}
M_{k=j}=\mu_0 r_{st} \left[ \ln{\frac{8r_{st}}{d_R}}-\frac{1}{2}+\frac{d_R^2}{32r_{st}^2} \left( \ln{\frac{8r_{st}}{d_R}}+\frac{1}{4} \right) \right]
\end{equation}

The  induced current $I^{ind}_j$ is calculated using Eq.\ref{eq:indmateq} and the calculated value of $F_j^{ext}$. The z-component of the induced magnetic flux density created by the current $I^{ind}_j=I^{ind}(z=z(j))$ of the $j^{th}$ ring is
\begin{eqnarray}
	B^{ring}_z(r_x=0,\Delta_z)&=&\frac{I^{ind}_j\mu_{0}}{4\pi}\int_{0}^{2\pi}\frac{R^{2}}{(R^{2}+\Delta_z^{2})^{\frac{3}{2}}}d\phi \nonumber\\
	&=&\frac{I^{ind}_j\mu_{0}}{2}\frac{R^{2}}{(R^{2}+\Delta_z^{2})^{\frac{3}{2}}},
\end{eqnarray}
where the radius is $R=r_{st}$, and the distance between the rings is $\Delta_z=\Delta_{kj}=|z^{st}_k-z^{st}_j|$.  The induced magnetic flux density along the axis of the shielding tube is determined by the summation over all rings
\begin{equation}
B^{ind}_{r_x=0}=\sum_{i=1}^{N_{st}}B^{ring}_z(\Delta_{kj}).
\end{equation}
The residual magnetic flux density is the superposition of the external and induced magnetic flux densities
\begin{equation}
B^{res}=B^{ext}+B^{ind}.
\end{equation}

\begin{figure}[h]
	\centering
\includegraphics[scale=0.6]{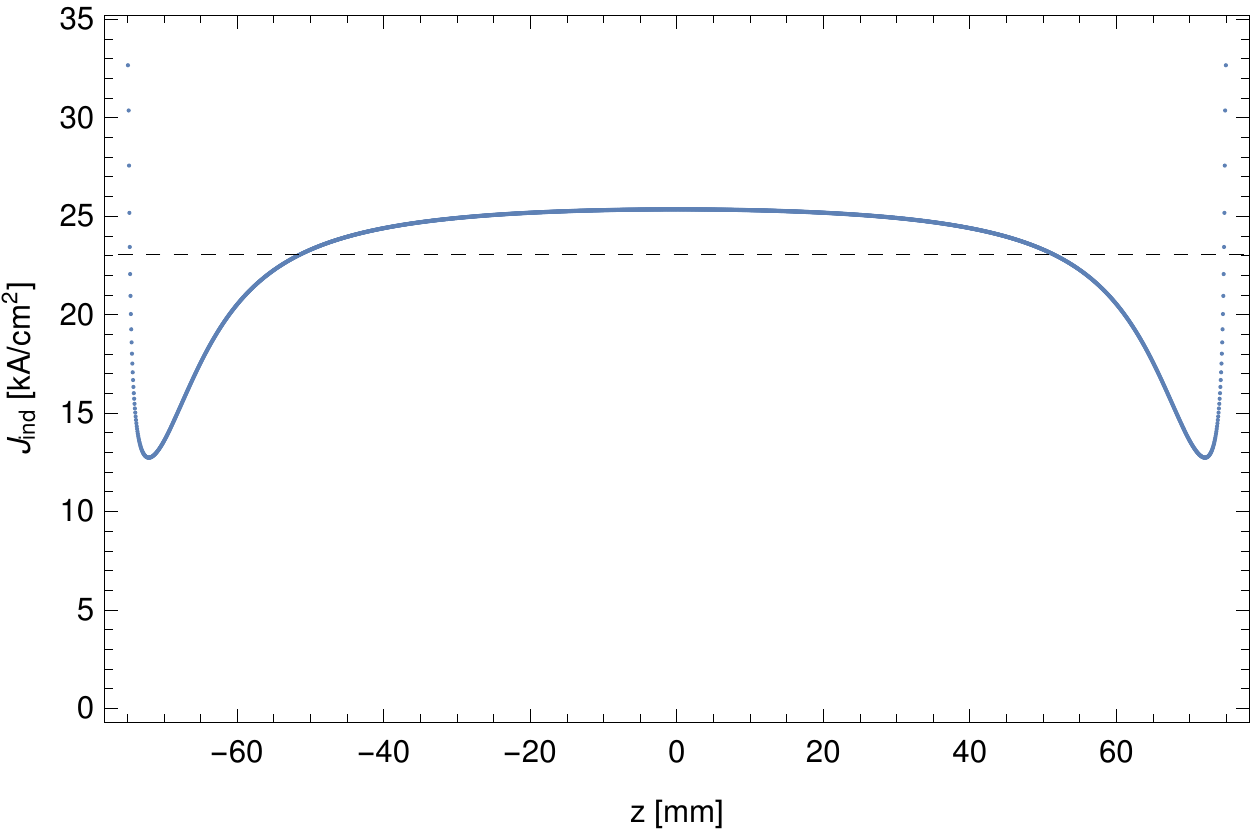}
	\caption{The simulated induced current density along the axis of the shielding tube. The dashed line indicates the average current density.}
	\label{fig:indcurr}
\end{figure}

The magnetic flux in the shielding tube of the external magnet  at $B_{ext}$=1~T, is calculated and used to calculate the induced current and induced current  density by solving Eq.~\ref{eq:indmateq}.  The calculated induced current density for $B_{ext}$=1~T is shown in Fig.~\ref{fig:indcurr}.

\begin{figure}[h]
\centering
    \includegraphics[scale=0.9]{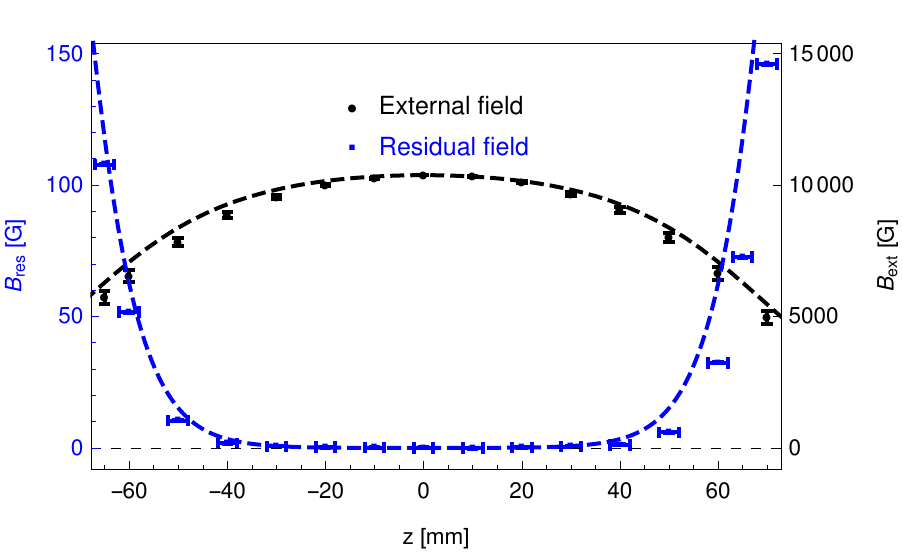}
	\caption{The measured values of $B_{res}$  (blue points) and $B_{ext}$ (black points).  The blue and black dashed curves represent the results of the simulations for $B_{res}$  and $B_{ext}$, respectively  (1~G=$10^{-4}$~T).}
	\label{comp}
\end{figure}

Figure~\ref{comp} shows the results of the calculations and the measured values on the external and residual magnetic field densities.   The data show a deviation from the simulated curve of the external field. This deviation is assumed to be  from the winding errors during the winding of the wire into the layers of a solenoid. The length of the holding structure need to be adjusted to the  diameter of the wire. The wire diameter alters in the range between 10~$\mu$m and 100~$\mu$m due to  the uncertainty of the wire coating during the fabrication of the wire and the gluing in the winding procedure. Therefore, the last winding at the end of a layer is not exactly a whole loop. This leads to an imperfection in the profile of the windings resulting in a stronger magnetic field towards the center where the windings are well arranged.

 The simulated and measured values of the residual  field are consistent  within the uncertainties  in most of the position points.  A small deviation is observed for $z$-values larger than 40~mm where the measured data points are slightly shifted on the $z$-axis with respect to the $z=0$ position. This is assumed to be due to the uncertainty  on the manual positioning of the sliding rod and the Hall-probe.  This systematic error is not considered in the analysis.    The distributions of the residual flux density for different lengths of the shielding tube and the external magnet can be found in Ref.~\cite{Frohlich:2017ald}.  The numerical stability of the calculation is shown by the robustness of the solution by  varying the geometrical parameters and the convergence for refinement of the discretization. It is also shown the homogeneity of  the  calculated $B_{res}$ in the $x-y$ plane  of the shielding tube.

\section{Summary}
\begin{table*}
\begin{tabular}{c c c}
\hline 
 External field & 1 T & 1.4 T \\ 
\hline
Shielding factor &  $32\times 10^{4}$ $(95 $\%$ C.L.)$ & $ (12 \pm  1)\times 10^3$ \\
Induced current density $J_{ind}$ [A/cm$^2$] & $(23 \pm 2) \times 10^3$ & $ (33 \pm 3)\times 10^3$\\
Shielded length [mm] (tube length 150 mm)& $80\pm2$ & $80\pm2$ \\
Residual field  after 9~h at 1~T and 14~h at 1.4~T & $ (0.0 \pm 2.4) \times 10 ^{-7}$~T & $(2.70 \pm  0.15)\times 10^{-4}$~T\\
Residual field  in the range of 20 to 80~h&$ (0.499 \pm  0.003) \times 10^{-4}$~T & \\ \hline
\end{tabular}
\caption{The results of the  measurements  for the shielding efficiency of the Bi-2212 tube at 4.2~K for applied axial fields of 1~T and 1.4~T. The tube length is \mbox{150 mm} and has a wall thickness of \mbox{3.5 mm}.}
\label{tab:conclusion}
\end{table*}

The challenging problem of operating a transversely polarized target in a longitudinal magnetic field is addressed. The work is motivated by the possible future upgrade of the \PANDA spectrometer with the  installation of a transversely polarized target.   An initial feasibility study for shielding the 2~T longitudinal field created by the \PANDA solenoid is presented. The shielding performance of a large melt cast Bi-2212 tube  in axial magnetic fields is tested at a temperature of 4.2~K. A dedicated apparatus, consisting of the shielding tube, two external magnets, a Hall probe, and a dewar filled with liquid helium is built and used for the measurements.  A data acquisition system is developed  to control and record  the power supply settings and to readout and store the collected data.

It is demonstrated experimentally that a magnetic flux density at the center of the tube of  ($10140\pm22$)~G  can be completely shielded,  resulting  a shielding factor better than $32\cdot10^4$.   The residual field is monitored for up to 4 days  showing a stable shielding operation.  In addition,  a large volume  within the Bi-2212 tube of 150~mm  length can be homogeneously shielded.  A residual field density less than 1~G  is measured over a  length of  80~mm at the center.   At 1.4~T,   $B_{res}$  is measured to be only 1.2~G and increases to  2.7~G after 14 hours of operation.   The values of the shielding factor  and the induced current density of the shielding tube, determined from the  measurements at 1~T and 1.4~T,  are summarized in Tab.~\ref{tab:conclusion}.  They show a high shielding performance of the Bi-2212 tube up to 1.4~T.

This feasibility study shows that a Bi-2212 tube, operated at 4.2~K, has the shielding features needed for  employing of a transversely polarized target in the presence of intense longitudinal fields.  The shielding factor of the Bi-2212 tube at 4.2~K and at an applied external field of 1~T is about two orders of magnitude better than what has been previously achieved at 10~K~\cite{Fagnard_2010}. High attenuation of the longitudinal magnetic field up to 1.4~T is proven.     Based on this study,  a dedicated prototype that fulfills the geometrical and the performance requirements of the \PANDA experiment can be designed.  The numerical calculations that are developed here and validated by the experimental tests can be used for this purpose.


\section*{Acknowledgments}
We acknowledge the support from the A2 Collaboration and the mechanical and the electronics workshops at the Institute for the Nuclear Physics (Mainz, Germany) in the preparation of the apparatus and the realization of the measurements. B.~F.  acknowledges useful and inspiring discussions with Christian Kremers from Dassault Syst\`emes  and Steffen Elschner from the University of Applied Science Mannheim. We thank Florian Feldbauer, Cristina Morales and  Patricia Aguar Bartolome  (Helmholtz-Institut Mainz) and Oleksandr Kostikov (Institute of Nuclear Physics, Mainz) for their help in this work and/or for taking shifts during the measurements.

\bibliography{mybibfile}

\end{document}